\documentclass[twocolumn,english,aps,graphicx,amsmath,showpacs]{revtex4}
\usepackage[T1]{fontenc}
\usepackage[latin1]{inputenc}
\usepackage{babel}
\usepackage{graphics}

\makeatletter

\providecommand{\LyX}{L\kern-.1667em\lower.25em\hbox{Y}\kern-.125emX\@}

\makeatother
\begin{document}

\title{Synchronization of Mutually Versus Unidirectionally Coupled Chaotic Semiconductor Lasers}

\author{Noam Gross\( ^{1} \), Wolfgang Kinzel\( ^{2} \), Ido Kanter\( ^{1} \), Michael Rosenbluh\( ^{1} \), Lev
Khaykovich\(^{1} \)}

\affiliation{\( ^{1} \)Department of Physics, Bar-Ilan University,
Ramat-Gan, 52900 Israel,}

\affiliation{\( ^{2} \)Institut f\"ur Theoretische Physik,
Universit\"at W\"urzburg, Am Hubland 97074 W\"urzburg, Germany}

\begin{abstract}
Synchronization dynamics of mutually coupled chaotic semiconductor
lasers are investigated experimentally and compared to identical
synchronization of unidirectionally coupled lasers. Mutual
coupling shows high quality synchronization in a broad range of
self-feedback and coupling strengths. It is found to be tolerant
to significant parameter mismatch which for unidirectional
coupling would result in loss of synchronization. The advantages
of mutual coupling are emphasized in light of its potential use in
chaos communications.
\end{abstract}




\maketitle Communication via chaotic signals has drawn much
attention in the last two decades. A semiconductor laser subjected
to an external feedback displays complex chaotic behavior
\cite{Tartwijk1995,Heil1998,Miles1980,Lenstra1985}. Two chaotic
lasers have been shown to synchronize with each other and are
excellent candidates for fast and secure
communication\cite{Mirasso2004,Buldu2004,Argyris2005}. Recently, a
field experiment using long fiber spans of commercial optical
networks has been conducted, in which chaotic optical
communications at high transmission and low bit-error rates were
reported \cite{Argyris2005}. In this experiment, a receiver laser
was synchronized to a transmitter laser unidirectionally allowing
unidirectional information flow only. The advantage of a mutually
synchronized system is that it increases the efficiency of the
apparatus by allowing a bilateral conversation. Moreover
unidirectional communication is a private-key system where the
system parameters serve as the secret key. In ref.
\cite{Klein2005} it was shown that it might be possible to use the
synchronization of two mutually coupled symmetric chaotic systems
in a novel cryptographic key-exchange protocol, whereby secret
messages can be transmitted over public channels without using any
previous secrets. Recently we made a first step toward realization
of such protocol by suggesting a mutual chaos pass filter
procedure based on experiments which revealed a window of
parameters where mutual coupling is advantageous over its
unidirectional counterpart \cite{Klein2006b}. Here we explore the
robustness of mutually coupled lasers to different experimental
parameters.

Chaos synchronization via unidirectional optical coupling is of
two types: identical and generalized
\cite{Buldu2004,Liu2003,Murakami2002,Vicente2002}. Identical, also
known as anticipated synchronization, appears when two nearly
identical lasers are subjected to the same optical feedback. One
laser simply reproduces the dynamics of the other. The most simple
and effective way to achieve this type of synchronization is by
setting the transmitter laser's ($TL$) external feedback strength,
$\kappa_{t}$, and the receiver laser's ($RL$) coupling strength,
$\sigma_{r}$, to be equal to each other, $\kappa_{t}=\sigma_{r}$,
while the $RL$ external feedback, $\kappa_{r}=0$ \cite{Buldu2004}.
Generalized synchronization requires strong injection (coupling
strength) and the $RL$ behaves more like a driven oscillator with
it's output driven by the injection signal \cite{Liu2001}. Such
synchronization is robust to channel disturbances \cite{Liu2003}
and works well even under non identical laser parameters, hence it
is favorable for chaos communication
\cite{Argyris2005,Mirasso2004,Vicente2002,Heil2002,Kanakidis2006}.
Although identical synchronization can achieve higher fidelity, it
is difficult to realize in real systems due to its high
sensitivity to laser parameter mismatch
\cite{Liu2003,Liu2001,Liu2002}.

Mutual optical coupling has been explored mostly in a face to face
configuration, in which the lasers have no external cavities, or
self feedback \cite{Heil2001,Mulet2004,Erzgraber2005}. In this
case, achronal synchronization was found, for which one of the
lasers is a leader and the other a laggard in a time dependent
manner, thus assigning asymmetric physical roles to the lasers,
even under symmetric operating conditions. Recently, we introduced
mutual optical coupling with the addition of self feedback and
observed isochronal synchronization \cite{Klein2006}, i.e. the
lasers are synchronized with zero time delay. A necessary
condition for this type of synchronization is that the self
feedback round-trip lengths of both lasers be equal to the
coupling length (the distance between them). When both lasers are
subjected to approximately the same feedback, high quality
identical synchronization is achieved.

In this paper we focus on isochronal synchronization of mutually
coupled lasers. We explore the feedback and coupling strength
phase-space and show that this type of identical synchronization
has much higher tolerance for parameter mismatch than its
unidirectional counterpart. Thus, not only does mutual coupling
allow a bidirectional flow of information, it also combines the
high correlation offered by identical synchronization with the
robustness typical of generalized synchronization.


\begin{figure}

{\centering \resizebox*{0.5\textwidth}{0.4\textheight}
{{\includegraphics{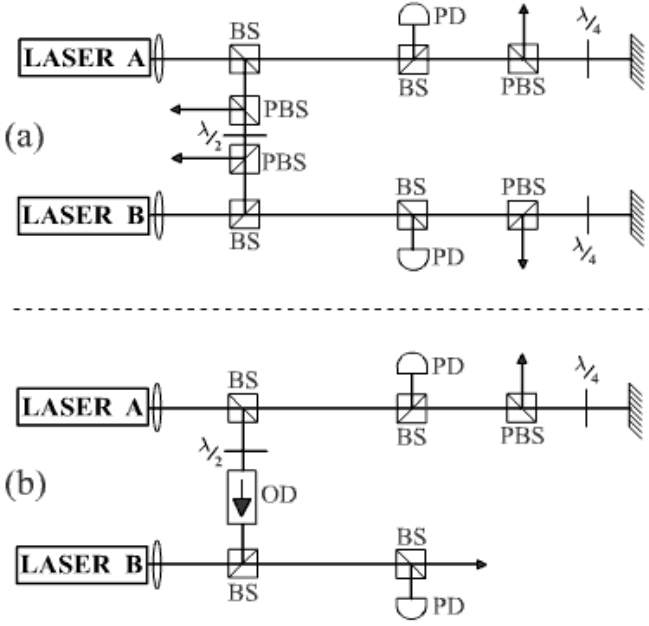}}}
\par}
\caption{\label{schema}Schematic of two coupled lasers for mutual
$(a)$ and unidirectional $(b)$ synchronization. BS - Beam
Splitter; PBS - Polarizing Beam Splitter; OD - Optical Isolator;
PD - Photodetector}
\end{figure}

Our experimental setup is depicted in Fig.\ref{schema}$(a)$ and in
Fig.\ref{schema}$(b)$ for mutual and unidirectional coupling,
respectively. We use two single-mode lasers emitting at 660 nm and
operating close to their threshold. The lasers are "off-the-shelf"
non-preselected devices, however, at room temperature they emit at
nearly identical optical wave length ($\Delta \lambda$ = 0.2nm)
and have the same threshold current (43.1mA) and P/I curve (better
than 99\% match). The temperature of each laser is stabilized to
better than 0.01K and the absolute temperature of each laser is
adjusted so as to match the individual laser wavelengths to be
identical. The length of the external cavities is set to 180 cm
(round trip time 12 ns). Self feedback strength is adjusted using
a $\lambda/$4 wave plate and a polarizing beam splitter. In the
mutual coupling experiment (Fig.\ref{schema}$(a)$), the two lasers
($A$ and $B$) are mutually coupled by injecting a fraction of each
one's output power to the other. Coupling power is adjusted using
a $\lambda/$2 wave plate and two polarizing beam splitters. In the
unidirectional coupling experiment (Fig.\ref{schema}$(b)$), laser
$B$ is coupled unidirectionally to laser $A$, with
unidirectionality ensured by an optical diode (-34 dB). Coupling
power is adjusted using a $\lambda/$2 wave plate located in front
of the optical isolator. In both cases, coupling optical paths are
set to be equal the self feedback round trip path. Two fast
photodetectors (response time $<$ 500 ps) are used to monitor the
laser intensities which are simultaneously recorded with a digital
oscilloscope (500MHz, 1GS/s).


\begin{figure}

{\centering \resizebox*{0.5\textwidth}{0.3\textheight}
{{\includegraphics{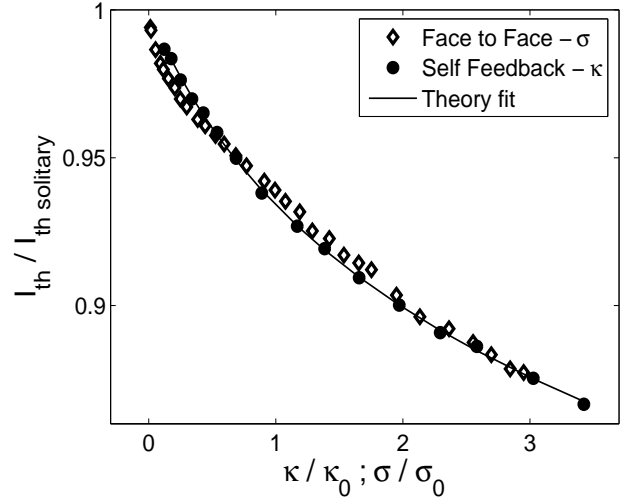}}}
\par}
\caption{\label{K_S_Ith}Measured feedback strength versus the
reduction in laser threshold current. $\sigma_{0}$ and
$\kappa_{0}$ are the feedback strength values required for a 6.6\%
reduction in $I^{sol}_{th}$. Theoretical fit is from $eq.(1)$ (see
text).}
\end{figure}

The feedback strength of a laser is measured via the reduction of
the laser's threshold current. The relation between the threshold
current, $I_{th}$, of a laser with external feedback and the
threshold current a solitary laser, $I^{sol}_{th}$, is
\cite{Olsson1984}

\begin{eqnarray}
\frac{I_{th}}{I^{sol}_{th}}= 1 - \gamma \ln(1+\delta
\frac{\kappa}{\kappa_{0}})
\end{eqnarray}
where $\gamma$ and $\delta$ are constants, $\kappa$ is the
feedback strength and $\kappa_{0}$ is the feedback strength
equivalent to a reduction of 6.6\% in $I^{sol}_{th}$. $\kappa$ is
proportional to the fraction of light power coupled back into the
cavity. In Fig.\ref{K_S_Ith} this relation is confirmed
experimentally (closed circles). The theoretical fit provides the
values $\gamma=0.0754$ and $\delta=1.40$. We also find that the
feedback strength between two coupled lasers behaves in agreement
with $eq.(1)$. To measure the influence of the coupling strength
on the laser's threshold current, the lasers were set in a face to
face configuration, in which they are exposed to mutual feedback
without having an external cavity of their own. Laser wavelengths
were set by temperature control to achieve a maximum overlap
between individual laser modes. This measurement is also shown in
Fig.\ref{K_S_Ith} as open diamonds. Similar to self feedback,
$\sigma_{0}$ is equivalent to a 6.6\% reduction in $I^{sol}_{th}$
of both lasers, and $\sigma$ is the coupling strength. We
attribute the small differences between the self and mutual
feedbacks due to the residual difference in lasers frequencies in
the case of mutual feedback.


\begin{figure}

{\centering \resizebox*{0.5\textwidth}{0.3\textheight}
{{\includegraphics{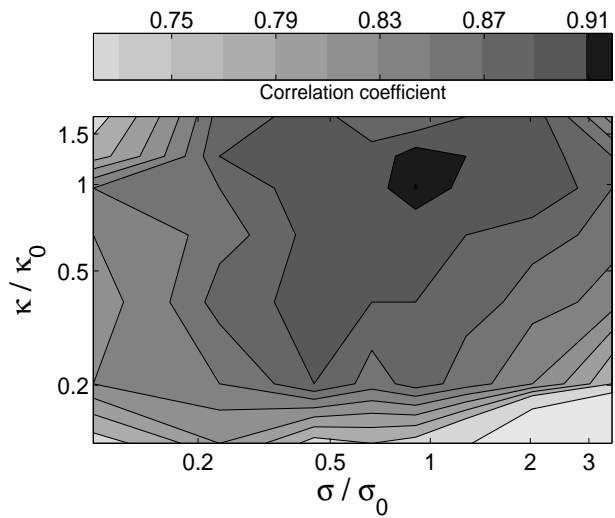}}}
\par}
\caption{\label{Mut_K_S}Mutual coupling phase space where the
correlation coefficient (gray level) is plotted as a function of
$\kappa$ and $\sigma$.}
\end{figure}

The phase-space of when good synchronization is achieved as a
function of self feedback ($\kappa$) and mutual coupling strength
($\sigma$) is shown in Fig.\ref{Mut_K_S}. At any point in this
phase space, symmetry between the two lasers is maintained so that
$\kappa=\kappa_{A}=\kappa_{B}$ and $\sigma=\sigma_{A}=\sigma_{B}$.
The quality of synchronization between lasers is evaluated by the
correlation coefficient \cite{Klein2006}, $\rho$ defined as
follows:

\begin{equation}\label{eqrho}
    \rho = \frac{\sum^{i}(I_{A}^{i}-<I_{A}^{i}>)\cdot{(I_{B}^{i}-<I_{B}^{i}>})}{\sqrt[]{\sum^{i}(I_{A}^{i}-<I_{A}^{i}>)^{2}\cdot\sum^{i}(I_{B}^{i}-<I_{B}^{i}>)^{2}}}
\end{equation}
where $I_{A}^{i}$ and $I_{B}^{i}$ are the instantaneous
intensities of lasers A and B respectively (see note
\cite{CorrelationNote}). The best synchronization ($\rho\geq0.91$)
is achieved for feedback values of $\kappa_{0}$ and $\sigma_{0}$.
The synchronization, however, is robust to large changes in
feedback strength and stable isochronal synchronization is
maintained even for a 50\% deviation from these values. This
robustness is very encouraging considering real communication
applications where channel intensity (and thus feedback strength)
might suffer from fluctuations and disturbances. A theoretical
phase-space, based on a numerical solution of the Lang-Kobayashi
equations is available in ref. \cite{Klein2006} and confirms the
existence of a broad area where the lasers exhibit stable
synchronization. When considering the case of $\kappa=\sigma$,
stable synchronization is intuitively understood since each laser
is subjected to a feedback consisting of equal contributions from
both laser intensities, hence their "driving force" is identical.
Our measurements, however, show that $\kappa=\sigma$ is not
necessary and the coupling intensity can be significantly weakened
without destroying the correlation, thus allowing for conditions
favorable for the realization of secure information exchange
\cite{Klein2005,Klein2006b}.

\begin{figure}

{\centering \resizebox*{0.5\textwidth}{0.3\textheight}
{{\includegraphics{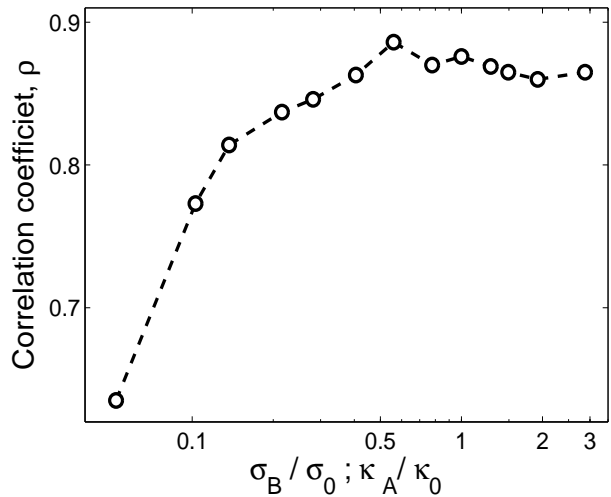}}}
\par}
\caption{\label{Uni_K_S}Correlation coefficient as a function of
$\kappa_{A}$ and $\sigma_{B}$ for unidirectional coupling.
$\kappa_{A}=\sigma_{B}$ in all cases.}
\end{figure}

The unidirectional coupling phase space is presented in
Fig.\ref{Uni_K_S}. Here, the $TL$ (laser $A$) is subjected only to
a self feedback from its external cavity ($\kappa_{A}$) and the
$RL$ (laser $B$) is subjected only to coupling from laser $A$
($\sigma_{B}$), while the condition of a total feedback strength
symmetry is maintained ($\kappa_{A}=\sigma_{B}$). As in mutual
coupling, good synchronization is achieved in the vicinity of
$\kappa_{0}$ and $\sigma_{0}$. For $\kappa_{A}\neq\sigma_{B}$ the
correlation drops rapidly and therefor we examined an additional
self feedback to the $RL$, also known as a closed loop scheme
\cite{Mirasso2004}. However, all attempts resulted in
deterioration of the synchronization. In particular, for the
closed loop case, synchronization is very sensitive to the phase
of the returning field, demanding a careful adjustment of the
external cavity length to sub wavelength levels
\cite{Mirasso2004,Peil2002}. We note that we have not observed a
phase dependence in the configuration of mutually coupled lasers.


\begin{figure}

{\centering \resizebox*{0.5\textwidth}{0.3\textheight}
{{\includegraphics{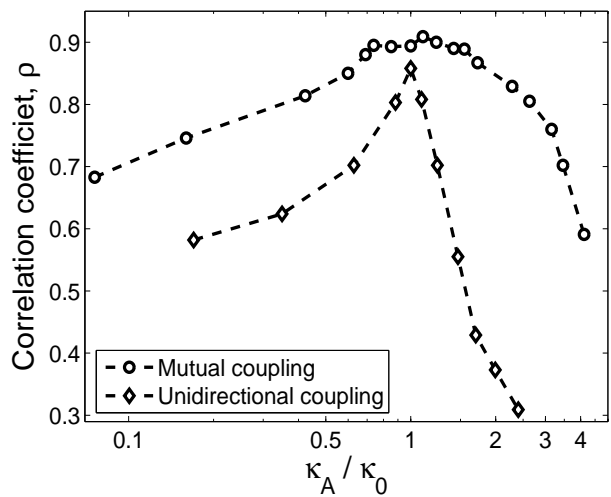}}}
\par}
\caption{\label{Kappa}Correlation coefficient for different values
of $\kappa_{A}$, laser $A$'s self feedback rate. Coupling strength
is $\sigma_{0}$. For mutual coupling (circles)
$\kappa_{b}=\kappa_{0}$ and $\kappa_{b}=0$ for unidirectional
coupling (diamonds).}
\end{figure}

In Fig.\ref{Kappa} we demonstrate the sensitivity to self feedback
strength for both unidirectional and mutual coupling
configurations. Correlation coefficient is calculated for
different values of $\kappa_{A}$, while $\kappa_{b}=\kappa_{0}$
for mutual coupling and $\kappa_{b}=0$  for unidirectional
coupling. In either cases Coupling strength is $\sigma_{0}$. For
unidirectional coupling (diamonds), good synchronization cannot be
attained unless $\kappa_{A}=\kappa_{0}$. However, for mutual
coupling (circles), small deviations of the self feedback strength
of one laser with respect to that of the other, does not affect
synchronization quality. In this case, good synchronization of
$\rho=0.9$ is obtained for self feedback range of
$0.7\kappa_{0}\leq\kappa_{A}\leq1.5\kappa_{0}$. Thus mutual
coupling exhibits high tolerance for asymmetry between the
operating lasers, while unidirectional coupling is very sensitive
to the mismatch and synchronization appears only if the condition
of total feedback symmetry is fulfilled.


\begin{figure}

{\centering \resizebox*{0.5\textwidth}{0.3\textheight}
{{\includegraphics{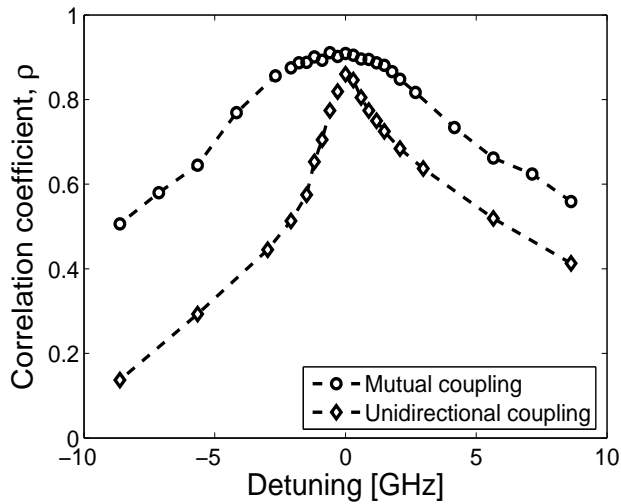}}}
\par}
\caption{\label{detuning}Correlation coefficient as a function of
detuning of laser $A$'s optical frequency with respect to laser
$B$, for mutual (circles) and unidirectional (diamonds) coupling.
Feedback strengths are $\sigma_{0}$ and $\kappa_{0}$.}
\end{figure}

We also examined the sensitivity of the two schemes to detuning of
the optical frequency of one laser with respect to the other. We
changed the temperature of laser $A$ while keeping laser $B$'s
temperature constant. Variation of the laser temperature shifts
the gain curve at a rate of 0.17 nm/K. More important, however, is
the shift of internal modes which we measured to be at a rate of
0.043 nm/K. In Fig.\ref{detuning}, the effect of detuning on
synchronization is shown. Unidirectional coupling (diamonds) is
very sensitive to the mismatch of the laser frequencies, as was
observed before in ref.\cite{Vicente2002,Liu2003}. Mutual
coupling, on the other hand, is robust and easily accommodates a
detuning of $\pm$ 2GHz ($\pm$ 0.07K), which corresponds to a
significant mismatch in the laser spectra \cite{BroadeningNote}.
The detuning of synchronized lasers results in a smaller spectral
overlap which effectively reduces the coupling strength $\sigma$.
For mutually coupled lasers the effective coupling can be varied
by a factor of 2 before a significant degradation in
synchronization quality is observed (see Fig.\ref{Mut_K_S}). For
unidirectional coupling, however, reduction of effective feedback
strength results in a sharp deterioration in the quality of
synchronization as shown in Fig.\ref{Kappa}.


To conclude, we have investigated experimentally the phase-space
of feedback and coupling strengths for two mutually coupled lasers
each subjected to an external feedback. We have found conditions
for which the parameter space for good synchronization is wide. A
comparison between unidirectional and mutual coupling, resulting
in identical synchronization, reveled that while the first is
extremely sensitive to deviations in the feedback strengths and
laser frequency detuning, the latter shows robust synchronization
even under non identical operation conditions and might be
suitable for optical communication applications \cite{Klein2006b}.



\begin{thebibliography}{9}

\bibitem{Tartwijk1995}G.H.M. van Tartwijk, A.M. Levine, and D. Lenstra, IEEE J. Sel. Top. Quantum
Electron. \textbf{1}, 466 (1995).

\bibitem{Heil1998}T. Heil, I. Fischer, and W. Els\"{a}{\ss}er,
 Phys. Rev. A \textbf{58}, R2672 (1998).

\bibitem{Miles1980}R.O. Miles, A. Dandridge, A.B. Tveten, H.F. Taylor, and T.G. Giallorenzi, Appl. Phys. Lett.
 \textbf{37}, 990 (1980).

\bibitem{Lenstra1985}D. Lenstra, B.H. Verbeek, and A.J. Den Boef, IEEE J. Quantum Electron. \textbf{QE-21}, 674 (1985).

\bibitem{Buldu2004}J.M. Buld\'{u}, J. Garc\'{i}a-Ojalvo, and M.C. Torrent, IEEE J. Quantum Electron. \textbf{40}, 640 (2004).

\bibitem{Argyris2005}A. Argyris, et al. Nature \textbf{438}, 343-346 (2005).

\bibitem{Mirasso2004}C.R. Mirasso, R. Vicente, P. Colet, J. Mulet, and T. P\'{e}rez,
C. R. Physique \textbf{5}, 613-622 (2004).

\bibitem{Klein2005}E. Klein, R. Mislovaty, I. Kanter and W. Kinzel, Phys.
Rev. E \textbf{72}, 016214  (2005).

\bibitem{Klein2006b}E. Klein, N. Gross, E. Kopelowitz, M. Rosenbluh,
W. Kinzel, L. Khaykovich and I. Kanter, cond-mat/0604569 (Yet
unpublished ).

\bibitem{Murakami2002}A. Murakami, and J. Ohtsubo, Phys. Rev. A \textbf{65}, 033826 (2002).

\bibitem{Vicente2002}R. Vicente, T. Perez, and C.R. Mirasso, IEEE J. Quantum Electron.
\textbf{38}, 1197 (2002).

\bibitem{Liu2003}Y. Liu, P. Davis, Y. Takiguchi, t. Aida, S. Saito, and J.M. Liu, IEEE J. Quantum Electron.
\textbf{39}, 269 (2003).

\bibitem{Liu2001}Y. Liu, H.F. Chen, J.M. Liu, P. Davis, and T. Aida, IEEE Transactions on Circuits and Systems-I
\textbf{48}, 1484 (2001).

\bibitem{Liu2002}Y. Liu, Y. Takiguchi, P. Davis, T. Aida, S. Saito, and J.M. Liu, Appl. Phys. Lett. \textbf{80},
 4306 (2002).

\bibitem{Heil2002}T. Heil, J. Mulet, I. Fischer, C.R. Mirasso, M. Peil, P. Colet, and W. Els\"{a}{\ss}er,
IEEE J. Quantum Electron. \textbf{38}, 1162 (2002).

\bibitem{Kanakidis2006}D. Kanakidis, A. Agryris, A. Bogris, and D. Syvridis, J. Lightwave Technology
 \textbf{24}, 335 (2006).

\bibitem{Heil2001} T. Heil ,I. Fisher, W. Els\"{a}sser, J. Mulet and
C.R. Mirasso, Phys. Rev. Lett. \textbf{86}, 795 (2001);

\bibitem{Mulet2004}J. Mulet, C. Mirasso, T. Heil, and I. Fischer, J. Opt. B: Quantum Semiclass. Opt.
\textbf{6}, 97-105 (2004).

\bibitem{Erzgraber2005}H. Erzgr\"{a}ber, D. Lenstra, B. Krauskopf, E. Wille, M. Peil, I. Fischer, and W. Els\"{a}{\ss}er,
 Opt. Comm. \textbf{255}, 286-296 (2005).

\bibitem{Klein2006}E. Klein, N. Gross, M. Rosenbluh,
W. Kinzel, L. Khaykovich and I. Kanter, cond-mat/0511648 (To
appear in Phys. Rev. E.)

\bibitem{Olsson1984}N.A. Olsson, and N.K. Dutta, Appl. Phys. Lett. \textbf{44}, 840 (1984).

\bibitem{CorrelationNote}We work in the well studied
LFF (Low Frequency Fluctuations) regime
\cite{Tartwijk1995,Heil1998,Miles1980,Lenstra1985}, characterized
by intensity breakdowns. These breakdowns reccur on a time scale
of hundreds of ns while the chaotic features we are interested in
are an order of magnitude smaller in intensity and fluctuate on
sub-ns time scales. In order to obtain accurate result for the
overlap, we divide the intensity traces into 10ns segments
(containing 10 sample points) and the correlation coefficient is
calculated between matching segments and then averaged over long
time stretches (including the intensity breakdowns themselves).

\bibitem{Peil2002}M. Peil, T. Heil, I. Fischer, and W. Els\"{a}{\ss}er, Phys. Rev. Lett. \textbf{88},
174101 (2002).

\bibitem{BroadeningNote}A chaotic semiconductor laser is known to execute a
spectral line broadening of a few GHz when subjected to an
external feedback \cite{Miles1980,Lenstra1985}.





\end{thebibliography}
\end{document}